\documentstyle[12pt]{article}

\def\rf#1{(\ref{eq:#1})}
\def\lab#1{\label{eq:#1}}
\def\nonu{\nonumber}
\def\br{\begin{eqnarray}}
\def\er{\end{eqnarray}}
\def\be{\begin{equation}}
\def\ee{\end{equation}}

\def\({\left(}
\def\){\right)}
\def\tr{\mathop{\rm tr}}                  
\newcommand{\dder}[2]{\frac{d{#1}}{d{#2}}}
\newcommand{\pder}[2]{\frac{\partial{#1}}{\partial{#2}}}
\def\a{\alpha}
\def\b{\beta}

\def\d{\delta}

\def\eps{\epsilon}

\def\g{\gamma}

\newcommand{\h}{\frac{1}{2}}

\def\om{\omega}

\def\p{\phi}

\def\pa{\partial}
\def\pr{\prime}

\def\t{\tau}

\def\ti{\tilde}

\newcommand{\ct}[1]{\cite{#1}}
\newcommand{\bi}[1]{\bibitem{#1}}
\newcommand {\1}[1]{\frac{1}{#1}}
\newcommand{\nit}{\noindent}
\newcommand\threemat[9]{\left(\begin{array}{ccc}  
{#1} & {#2} & {#3} \\ {#4} & {#5} & {#6} \\
{#7} & {#8} & {#9} \end{array} \right)}
 \newcommand{\sect}[1]{\setcounter{equation}{0}\section{#1}}
 
 \relax

\begin{document}

\nit
{\Large {\bf 
Darboux-Egoroff Metrics, Rational
Landau-Ginzburg Potentials and the Painlev\'e VI  Equation
}}

\bigskip
{\renewcommand{\thefootnote}{\fnsymbol{footnote}}
{\large\bf H. Aratyn$^1$\footnote{
     E-mail: {\tt aratyn@uic.edu}},
     J.F. Gomes$^2$\footnote{
     E-mail: {\tt jfg@ift.unesp.br}}
and A.H. Zimerman$^2$\footnote{
     E-mail: {\tt zimerman@ift.unesp.br}}
}}

{\sl
$^1$ Department of Physics,
University of Illinois at Chicago,
845 W. Taylor St.,
Chicago, IL 60607-7059 

$^2$ Instituto de F\'{\i}sica Te\'{o}rica-UNESP,
Rua Pamplona 145,
01405-900 S\~{a}o Paulo, Brazil
}

\begin{center}
{\bf Abstract}
\end{center}

We present a class of three-dimensional integrable structures 
associated with the Darboux-Egoroff metric and classical Euler 
equations of free rotations of a rigid body. They are obtained 
as  canonical structures of rational Landau-Ginzburg potentials
and provide solutions to the Painlev\'e VI equation.

\sect{Introduction}

The flat coordinates and prepotentials play a prominent role
in the study and classification of solutions 
to the Witten-Dijkgraaf-Verlinde-Verlinde (WDVV) equations 
\ct{witten,dvv,Du-lectures,Du3a}.
Our interest is in finding the formalism which would provide a universal
approach to canonical integrable structures behind the WDVV equations.
Here, we understand canonical integrable structures as hierarchies
formulated in terms of canonical coordinates and with
evolution flows obtained in a conventional way 
from the Riemann-Hilbert factorization problem with 
twisting condition imposed \ct{vandeLeur:2000gk,agz01}.
This approach coincides with the framework of Darboux-Egoroff
metric constructed in terms of the Lam\'e coefficients and their symmetric
rotation coefficients \ct{Du-lectures}.
Additional reduction (scaling invariance) effectively turns  the 
evolution flows into the isomonodromic deformations (up to the Laplace
transformation) and allows the flow evolution equations
to be rewritten as Schlesinger equations \ct{Du-lectures}.
One can associate a tau function to the factorization problem which, 
upon imposition of the conformal condition, coincides 
with the corresponding isomonodromic tau function \ct{LM,needs-paper}.
The tau function appears to be a key object in this construction due
to the fact that
all the other objects of canonical formalism (e.g. rotation 
coefficients) can be obtained from it.
The conformal condition introduces a notion of 
scaling dimension or homogeneity of the tau function which is convenient to
use in order to classify integrable models.
For the integer scaling dimensions of the tau function 
the integrable hierarchies can be reproduced by the (C)KP-like formalism 
\ct{Aratyn:2001cj,needs-paper}. The tau functions can be calculated
in the Grassmannian approach to the multi-component KP hierarchy 
\ct{LM,prepare}.
The structures with the tau-functions possessing fractional
scaling dimension lack such a clear underlying foundation.
An exception to this rule is the trivial two-dimensional
case where the tau function can be explicitly found 
for all models.

This work studies three-dimensional models
with the tau functions having the 
scaling dimension equal to the fractional number of a 1/4
in the hope of providing data for its future general formalism.
The integrable structures are obtained from
Frobenius manifolds associated with a class of rational 
Lax functions \ct{aoyama}. The construction of canonical coordinates
for these polynomials generalizes the well-known construction 
of canonical coordinates associated with the monic 
polynomials \ct{Du-lectures,Hlectures}.
It is worth noting that
the space of monic polynomials endowed with a natural metric 
provides one of the most standard examples of the Frobenius manifolds.

We find a closed expression for the tau function for
one of the considered models.
We also associate the canonical structure to algebraic solutions
for the Painlev\'e VI equations considered by Hitchin \ct{H1,H2,H3}
and Segert \ct{segert1,segert2}.

The paper is organized as follows.
In Section \ref{de-metric}, we define the 
canonical integrable structure behind the WDVV equations
emphasizing its connection to the Darboux-Egoroff metric.
The tau function plays a key role in this formalism.
In this section we also introduce 
homogeneity as a scaling dimension.
In addition, we discuss the relation of canonical integrable structures
to the flat coordinates, structure
constants and associativity equations.

The special case of three dimensions is presented in Section 
\ref{nthree} where the Darboux-Egoroff equations are shown to take 
the form of the  classical Euler 
equations of free rotations of a rigid body.
Also, the scaling dimension of the tau function is found
to be related to the integral of the  Euler 
equations.

The rational potentials and its metric along with the associated 
canonical coordinates related to the Darboux-Egoroff metric 
are described in Section \ref{lg-models}. 
Examples of the three-dimensional canonical integrable
models derived from the rational potentials are given
in Section \ref{examples} and its relation to the Painlev\'e VI
equation is established. The Darboux-Egoroff metric 
is described in detail for the discussed three-dimensional models.
The explicit form of the tau function 
is found for one of the models.
Section \ref{outlook} presents concluding remarks.

\sect{From factorization problem to Darboux-Egoroff metric}
\label{de-metric}

Let the ``bare'' wave (matrix) function be
defined in terms of canonical coordinates ${\bf u}=
(u_1, {\ldots} ,u_N)$ as
\begin{equation}
\Psi_0 ({\bf u}, z) = \exp\left(z {\sum_{j=1}^{N} 
E_{jj}u_j}\right)
= e^{z U}
\lab{dfirst}
\end{equation}
where $U= {\rm diag} \( u_1, {\ldots} ,u_N\)$
and the unit matrix $E_{jj}$ has the matrix elements 
$(E_{jj})_{kl}=\d_{kj} \d_{lj}$.

Define  the ``dressed''  wave matrix $\Psi$:
\be
\Psi ({\bf u}, z) = \Theta ({\bf u}, z)\, \Psi_0 ({\bf u}, z)  
\lab{gms}
\ee
obtained by acting on the ``bare'' wave (matrix) function 
$\Psi_0 ({\bf u}, z)$ with the dressing matrix :
\be
\Theta ({\bf u}, z)=
1+ \theta^{(-1)}({\bf u}) z^{-1}+ \theta^{(-2)} z^{-2}+ \cdots \, .
\lab{rwf}
\ee
The ``dressed''  wave matrix $\Psi$
enters a factorization problem  :
\be
\Psi ({\bf u}, z) g  (z) \,=
\Theta ({\bf u}, z)\Psi_0 ({\bf u}, z) g  (z) \,
= \,  M  ({\bf u},z) 
\lab{rh-def}
\ee
where $g$ is a map from $z \in S^1$ to the Lie group $G$
and $M  ({\bf u},z) $ is a positive power series in $z$ :
\be
M  ({\bf u},z) = M_0  ({\bf u}) + M_1  ({\bf u}) z+ M_2  ({\bf u}) z^2
+{\cdots} \,.
\lab{mpos}
\ee
If the map $g(z)$ is such that it satisfies the twisting condition
$ g^{-1} (z) = g^T (-z)$ then it follows that the matrices
$\Theta ({\bf u}, z)$ and $M  ({\bf u},z)$ from equation 
\rf{rh-def} will also satisfy the twisting conditions
$\Theta^T ({\bf u},- z)= \Theta^{-1}  ({\bf u}, z)$,
$M^T ({\bf u},- z)= M^{-1}  ({\bf u}, z)$ \ct{vandeLeur:2000gk,agz01}.
In this case, the matrix $\theta^{(-1)}({\bf u}) $ from expansion \rf{rwf}
must be symmetric,
$\theta^{(-1)}({\bf u}) = \( \theta^{(-1)}\)^T ({\bf u}) $
and the matrix $M_0  ({\bf u})$ from expression \rf{mpos}
must be orthogonal,
$M_0^{-1}  ({\bf u})=M_0^T  ({\bf u})$.

The factorization problem \rf{rh-def} leads to the flow equations:
\br
\frac{\partial}{\partial u_j} \Theta ({\bf u}, z) 
&=& - \left(\Theta z E_{jj} \Theta^{-1} 
\right)_{-} \Theta ({\bf u}, z) 
\lab{uthpos}\\
\frac{\partial}{\partial u_j} M ({\bf u}, z) 
&=&  \left(\Theta z E_{jj} \Theta^{-1} 
\right)_{+} M ({\bf u}, z) 
\lab{ummpos}
\er
where $({\ldots} )_{\pm}$ denote the projections on positive/negative
powers of $z$.
Due to equation \rf{uthpos} the one-form 
${\rm Res}_{z} \(\tr \(  {\Theta}^{-1}  z ({d  \Theta}/d z) dU \)\)$
is closed. Conventionally, one associates to this one-form 
a $\tau$-function through
\be
d \log \t = 
{\rm Res}_{z} \(\tr \(  {\Theta}^{-1}  
z \dder{\Theta}{ z} d U  \)  \) 
\lab{jnj-tau}
\ee 
where $d=\sum_{j=1}^N \pa_j d u_j$ with $\pa_j = \pa / \pa u_j$.
One can verify that this definition of the tau function
implies relation $\theta^{(-1)}_{ii}=-\pa_i \log \t$ for the 
diagonal elements of the $\theta^{(-1)}$-matrix.
We will parametrize the $\theta^{(-1)}$-matrix as follows
\be
\theta^{(-1)}_{ij} = \left\{ \begin{array}{cc} 
\b_{ij} & i \ne j \\
- \pa_i \log \t & i = j \end{array} \right.
\lab{ths}
\ee
Due to equation \rf{uthpos} 
the symmetric off-diagonal components $\beta_{ij}$ satisfy
\be
\pa_{j}  \b_{ik} = \b_{ij}\b_{jk}, \;\;
 \quad  \quad i\, ,j \, ,k\, \;{\rm distinct}
\lab{pajthik}
\ee

It follows from the definition \rf{gms} that the ``dressed''  wave matrix 
$\Psi$
will satisfy the flow equations identical to those
in \rf{ummpos} :
\be
\frac{\partial}{\partial u_j} \Psi ({\bf u}, z) 
=  \left(\Theta z E_{jj} \Theta^{-1} 
\right)_{+} \Psi ({\bf u}, z) 
\lab{upsipos}
\ee
The projection in \rf{upsipos} contains only two
terms : $\( \Theta z E_{jj} \Theta^{-1} \)_{+}=z  E_{jj} + V_j$
where $V_j \equiv \lbrack \theta^{(-1)} \, , \, E_{jj}\rbrack $.
Accordingly,
\be
\frac{\pa \Psi}{\pa u_j} =(z E_{jj}+V_j) \Psi 
\lab{cdjpsi}
\ee
and ${\pa \Theta}/{\pa u_j} =z \lbrack E_{jj} \, ,\, \Theta \rbrack
+V_j \Theta $. Projecting on diagonal directions yields
$\pa_i\pa_j \log \t= - \b_{ij}^2$ for $i\ne j$. Hence
the dressing matrix can be expressed in terms of one object
only, 
the $\t$-function.

We  now define Euler and identity vectorfields $E$ and $I$ as:
\be
E = \sum_{i=1}^N u_i \frac{\pa}{ \pa u_i} \;\;\;, \;\;\;
I = \sum_{i=1}^N  \frac{\pa}{ \pa u_i} 
\lab{euler}
\ee
Their action on  $\Psi$ can easily be found from \rf{cdjpsi}
by summing over $j$ :
\be
E (\Psi)= \( z U +V \) \Psi, \qquad I (\Psi)= z \Psi
\lab{eonpsi}
\ee
where we introduced 
matrix $V = \lbrack \theta^{(-1)} \, , \, U \rbrack$
with components
\be
V_{ij} = (u_j-u_i) \theta^{(-1)}_{ij} = (u_j-u_i) \beta_{ij}, \quad
i,j=1, {\ldots} ,N\, .
\lab{v-comps}
\ee
Similarly, $E (\Theta) = z\lbrack U \, ,\, \Theta \rbrack
+V \Theta $.

Notice, that $E (\Psi_0) = z d \Psi_0 /d z$.
We will require that the same equality holds for 
the action of the grading operator
$z d/dz$ and of the Euler operator $E$ on $\Psi$ :
\be
z \frac{d }{d z} \Psi= E (\Psi)= \( z U +V \) \Psi
\lab{cnf}
\ee
The relation \rf{cnf} is called the conformal condition 
\ct{Du-lectures}.
It holds provided $ E (\Theta) = z d \Theta /d z$.
Also, from equation \rf{uthpos} we get by summing over all indices 
$j$ that $I (\Theta)=0$. In particular, for $\theta^{(-1)}$ we get
\br
E (\b_{ij}) &=& - \b_{ij},  \quad I (\b_{ij})=0 \lab{ebib}\\
E (\pa_i \log \t) &=& - \pa_i \log \t,  \quad I (\pa_i \log \t)=0 
\lab{etit}
\er
Since, $ E \pa_i= \pa_i E - \pa_i$ we see that $E \log \t={\rm constant}$ 
or $ E (\t)  = {\rm constant}\; \t $ 
is compatible with the conformal condition \rf{etit}. This constant defines
the scaling dimension (or homogeneity) of the tau function.

Plugging relation $ E \Theta = z d \Theta /d z$ into the formula
\rf{jnj-tau} for the $\t$-function one obtains \ct{Aratyn:2001cj}:
\be
\pa_j \log \t = 
{\rm Res}_{z} \(\tr \(  {\Theta}^{-1}  
E \Theta E_{jj}   \)  \) = \h \tr \( V_j V\)
\lab{tauvjv}
\ee 
Comparing with expression for the isomonodromy tau-function $\tau_I$
\ct{Du-lectures} one concludes that $\tau_I={1}/{\sqrt \tau}$
\ct{LM,needs-paper}.
The isomonodromy tau-function $\tau_I$ gives the Hamiltonian formulation 
$\pa_j V = \{ H_j \, , \, V\}$ 
for equation 
\be
\pa_j V = \lbrack V_j  \, , \, V\rbrack 
\lab{pjv}
\ee
via the formula: $\pa_j \log \t_I = H_j$.
Equation \rf{pjv} follows from the 
compatibility of equations \rf{cdjpsi}, \rf{eonpsi} and \rf{cnf}.
One clearly sees from \rf{tauvjv} and \rf{pjv} that
\be
I (\log \t) = 0 , \quad \;I(V)=0
\lab{ilogtv}
\ee
since $\sum_{j=1}^N V_j=0$.

The similarity transformation $V \to 
{\cal V} = M_0^{-1} V M_0 $ transforms $V$ to the constant matrix
${\cal V}$ ($\pa_j {\cal V} =0$) due to the flow equations $\pa_j M_0=
V_j M_0$, which follow from \rf{ummpos}.
Assume, now that there exists an invertible 
matrix $S$ which diagonalizes ${\cal V}$ :
\be
S^{-1} {\cal V}S=\mu = \sum_{j=1}^N \mu_j E_{jj} 
\lab{mudef}
\ee
where $\mu$ is the constant diagonal matrix .
Next, define a matrix
\be
M(u)=M_0(u)S=(m_{ij}(u))_{1\le i,j\le N},
\lab{mdef}
\ee
which governs transformation from the canonical coordinates $u_j,
j=1,{\ldots} ,N$ to the flat coordinates $x^\a, \a=1,{\ldots} , N$.
Let, constant non-degenerate metric be given by the matrix :
\be
\eta=(\eta_{\alpha\beta})_{1\le \alpha,\beta\le N}=M^TM=S^TS,
\quad \hbox{and denote}\quad \eta^{-1}=(\eta^{\alpha\beta})_{1\le \alpha,\beta\le N},
\ee
hence $\eta_{\a \b} 
= \sum_{i=1}^N m_{i\a} m_{i\b}$. 
Then 
the derivatives with respect to the flat coordinates $x^{\a},\, \a=1,{\ldots} 
,N$ are given by
\be
\frac{\pa }{\pa x^{\a}} = 
\sum_{i=1}^N \frac{m_{i\a}}{m_{i1}} \frac{\pa }{\pa u_i} 
\lab{flatt}
\ee 
with the reversed relation being
\be
\frac{\partial }{\partial u_i}
=\sum_{\alpha,\beta=1}^N 
\eta^{\alpha\beta}m_{i1}m_{i\beta}\frac{\pa }{\pa x^{\a}}.
\ee
The structure constants 
\be
c_{\a \b \g} = \sum_{i=1}^N
\frac{m_{i\a}m_{i\b}m_{i\g}}{m_{i1}}.
\lab{cstrus}
\ee
satisfy the associativity equation 
\be
\sum_{\d \g=1}^N c_{\a\b\d} \eta^{\d \g} c_{\g\sigma \rho} 
= \sum_{\d \g=1}^N c_{\a\sigma\d}\eta^{\d \g}
c_{\g\b\rho} \quad \; ,\; \a, \b, \sigma =1 ,{\ldots} ,N
\lab{wdvv}
\ee
and are given by derivations of the prepotential $F$ :
\be
c_{\a \b \g} = \frac{\pa^3 F}{\pa x^\a \pa x^\b \pa x^\g}\, . 
\lab{cfxx}
\ee
The metric $g=\sum_{\alpha,\beta=1}^N \eta_{\a \b}
dx^\alpha dx^\beta$ equals in terms of the canonical coordinates to :
\be
g=\sum_{i=1}^N h_i^2(du_i)^2
\lab{heta}
\ee 
with Lam\'e coefficients $h_i=m_{i1}$ being such that the corresponding
rotation coefficients :
\be
\b_{ij} = \frac{1}{h_j}\frac{\pa h_i}{\pa u_j}
\lab{rotco}
\ee
satisfy conditions of the Darboux-Egoroff metric, 
namely $\b_{ij} = \b_{ji}$ together
with the relations \rf{pajthik} and $ I(\b_{ij})=0$.
Note, that the Darboux-Egoroff condition $\b_{ij} = \b_{ji}$
is equivalent to square of the Lam\'e coefficient being a gradient
of some potential $\p$ : $h_i^2= \pa \p / \pa u_i$.

Furthermore, the Euler operator is given in terms of the flat coordinates as:
\be 
E = \sum_{\a=1}^N \( d_\a x^\a + r_\a \) \frac{\pa}{\pa x^\a}
\lab{eflat}
\ee
with $d_\a r_\a =0 $ and $d_\a= 1+\mu_1 -\mu_\a$ \ct{Du-lectures}.
The quasi-homogeneity condition states that :
\be
E (F) = d_F F + \, {\rm quadratic~terms}
\lab{quasi}
\ee
where the number $d_F$ denotes the degree of the prepotential $F$.

Following \ct{Du-lectures,segert2}, we will say that a function $\psi$ is of 
"homogeneity c" or ``scaling dimension c'' or ``$E$-degree'', if
$E (\psi) = c \, \psi$ for a constant $c$.

For the Lam\'e coefficients $h_i$ with $\sum_{i=1}^N h_i^2
=\sum_{i=1}^N m_{i1}^2 = \eta_{11}$ the homogeneity must be zero
for the constant metric tensor with $\eta_{11}\ne 0$ since
\be
0=E (\sum_{i=1}^N h_i^2)= \sum_{i=1}^N 2 h_i c h_i= 2 \eta_{11} c
\lab{homoghi}
\ee
{}From relation $E(\eta_{\a\b})= (d_F-d_1-d_\a-d_\b) \eta_{\a \b}$,
obtained by acting with the Euler vectorfield on both sides of
\rf{cfxx}, it follows that $d_F=3d_1$ for $\eta_{11}\ne 0$ 
and if $d_F\ne 3d_1$ then $\eta_{11}= 0$.
Hence the value of the homogeneity of the prepotential 
indicates when the homogeneity of the Lam\'e coefficients vanish.

For the class of models we consider here the 
Lam\'e coefficients $h_i$ are given by the formula:
\be
h_i^2= \pder{x_1}{u_i} \lab{hix1}
\ee
which agrees with a general feature of Frobenius manifolds 
endowed with the invariant metric \ct{Du-lectures}.
Accordingly, the homogeneity of the Lam\'e coefficients is a constant
number equal to $(\sum_{\a=1}^N d_\a \eta^{\a 1} -1 )$ according
to 
\be
E (h_i^2) = E \pa_i (x_1) = \pa_i E (x_1) - \pa_i  (x_1)
= \(\sum_{\a=1}^N d_\a \eta^{\a 1} -1 \) h_i^2
\lab{lamhom}
\ee
Hence, for $\eta_{11}=0$ it holds that $\sum_{\a=1}^N d_\a \eta^{\a 1}=1$.

Note, also that $E (h_i) = c h_i$ is consistent with 
the conformal condition
in \rf{ebib} for the arbitrary constant homogeneity $c$.

For the Darboux-Egoroff metric the identity vectorfield
vanishes when acting on the Lam\'e coefficient
as follows from
\be
I (h_i)= \sum_{j=1}^N \b_{ij}h_j= \sum_{j=1}^N  
\frac{1}{h_i}\frac{\pa h_j}{\pa u_i} h_j
= \frac{1}{2h_i} \frac{\pa \eta_{11}}{\pa u_i}=0
\lab{ihj}
\ee

\sect{The Three-dimensional Case}
\label{nthree}

Let us now consider the three-dimensional manifolds.
In this case, we can rewrite the antisymmetric matrix $V$ as:
\be
V= \threemat{0}{\om_3}{-\om_2}{-\om_3}{0}{\om_1}{\om_2}{-\om_1}{0}
\lab{vome}
\ee
or $(V)_{ij}= (u_j-u_i) \beta_{ij} =\eps_{ijk} \om_k$.
{}From \rf{ebib} and \rf{ilogtv} we see
that $\omega_k$ vanishes when acted on 
by the vectorfields $E$ and $I$. That  makes  $\omega_k$ 
effectively a function of one variable $s$ such that 
$E (s)= I (s)=0$.
Let us choose 
\be
s= \frac{u_2-u_1}{u_3-u_1}
\lab{sdef}
\ee
Then equation \rf{pjv} takes a form equivalent to
the Euler top equations:
\br
\frac{d \om_1}{d s}&=& \frac{\om_2\om_3}{s} \lab{euta}\\
\frac{d \om_2}{d s}&=&  \frac{\om_1\om_3}{s(s-1)} \lab{eutb}\\
\frac{d \om_3}{d s}&=& \frac{\om_1\om_2}{1-s} \lab{eutc}
\er
One verifies that $ d (\sum_{k=1}^3 \om^2_k)/d s =0$.
Consequently, 
\be
 \sum_{k=1}^3 \om^2_k = - R^2
\lab{rconst}
\ee
with $R$ being a constant is the integral of equations \rf{euta}-\rf{eutc}.
The same constant $R$ characterizes the 
homogeneity of the tau function, as we will show now.
Starting from expression \rf{tauvjv} one finds
for the scaling dimension \ct{Dubrovin:2001}
\be
E (\log \tau) = \h \sum_{j=1}^3 u_j\tr \( V_j V\)
= \h \tr \(  V^2\) = \h \tr \(  \mu^2\) = \h \sum_{\a=1}^3 \mu_\a^2
\lab{wtau}
\ee
Recalling that $(V)_{ij}= \eps_{ijk} \om_k$
we can rewrite the above as :
\be
E (\log \tau) = \h \sum_{j=1}^3 \sum_{i=1}^3 (\eps_{ijk} \om_k)^2
= - \sum_{k=1}^3 \om^2_k = R^2
\lab{etaur}
\ee
and since $\mu_\a= 1-d_\a +d/2$ with $d= d_F-3$
\be
R^2 =  \sum_{\a=1}^3 (\h -d_F/2+d_\a)^2
\lab{randdf}
\ee
We have seen above, that for $\eta_{11}$ different from zero the
homogeneity of the Lam\'e coefficients $h_i$ must vanish.
In such case, the Lam\'e coefficients $h_i$ depend only on one
variable $s$ due
to the fact that $I(h_i)=E(h_i)=0$.
The relations $\pa_j h_i^2 = \pa_i h_j^2$ translate for the function
$h_i^2(s)$ to
\be
s \dder{h_1^2}{s} =(s-1)s \dder{h_2^2}{s} =(1-s) \dder{h_3^2}{s} 
\lab{hofs}
\ee
Also, since
\be
\om_k = \frac{u_j-u_i}{2 h_ih_j} \pder{h_i^2}{u_j}
, \quad i,j,k \;\mbox{cyclic}
\lab{omkhij}
\ee
we find e.g.
\be
\om_3 =  \frac{s}{2h_1h_2} \dder{h_1^2}{s}, \;\;
\om_2 = \frac{s}{2h_1h_3} \dder{h_1^2}{s}
\ee
and so $h_3 \om_2=h_2 \om_3$ and similarly $h_1 \om_2=h_2 \om_1$.
We conclude that
\be
\om_i^2 = -\frac{R^2}{\eta_{11}} h_i^2, \quad i=1,2,3
\lab{omihi}
\ee
and comparing equations \rf{euta}-\rf{eutc} with equation \rf{hofs}
we obtain like in \ct{segert2} :
\be
s \dder{h_1^2}{s} =(s-1)s \dder{h_2^2}{s} =(1-s) \dder{h_3^2}{s} 
= -2 i \frac{R}{\sqrt{\eta_{11}}} h_1 h_2 h_3
\lab{hofsa}
\ee

\sect{Rational Landau - Ginsburg Models }
\label{lg-models}
Following Aoyama and Kodama \ct{aoyama} we study a rational potential :
\br
W (z) &=& \frac{1}{n+1} z^{n+1} +a_{n-1} z^{n-1}+ {\ldots} +a_0
+ \frac{v_1}{z-v_{m+1}}+ \frac{v_2}{2(z-v_{m+1})^2} + {\ldots} 
\nonu\\
&+&
\frac{v_m}{m(z-v_{m+1})^m}
\lab{ratW}
\er
which is known to characterize
the topological Landau-Ginzburg (LG) theory.
The rational potential in this form can be regarded as the Lax
operator of a particular reduction of the dispersionless 
KP hierarchy \ct{aoyama,Chang:2001qz,strachan}

The space of rational potentials from
 \rf{ratW} is naturally endowed with the metric :
\be
g ( \pa_t W, \pa_{t^{\pr}} W) = {\rm Res}_{z \in {\rm Ker} W^{\pr}}
\( \frac{ \pa_t W \pa_{t^{\pr}} W}{W^{\pr}} \)dz
\lab{wmetric}
\ee
where $\pa_t W = \pa_t a_{n-1} z^{n-1}+ {\ldots} + \pa_t a_0
+ \frac{\pa_t v_1}{z-v_{m+1}} +{\ldots} $
describes a tangent vector to the space of rational potentials obtained by
taking derivative of all coefficients with respect to their argument.
${W^{\pr}} (z)$ is a derivative with respect to $z$
of the  rational potential $W$ :
\be
{W^{\pr}} (z)=  z^{n} +(n-1) a_{n-1} z^{n-2}+ {\ldots} 
- \frac{v_m}{(z-v_{m+1})^{m+1}} 
\lab{wprime}
\ee
Next, we find the flat coordinates $x_\a, \a=1,{\ldots} ,m+1$
and ${\ti x}_\g, \g =1 ,..,n$ such that 
\be
g (\frac{\pa W}{ \pa x_\a}, \frac{\pa W }{\pa x_\b})
= \eta_{\a\b}, \;\;
g (\frac{\pa W }{ \pa {\ti x}_\g}, \frac{\pa W }{\pa {\ti x}_\d})
= {\ti \eta}_{\g\d},\;\;
g (\frac{\pa W }{ \pa x_\a}, \frac{\pa W }{\pa {\ti x}_\g})
= 0
\ee
with constant and non-degenerate matrices $\eta_{\a\b}$ and ${\ti \eta}_{\g\d}$.

Consider first the function $w=w(W,z)$ such that $W(z) = w^{-m}/m$ 
and $z = x_{m+1} +x_{m} w + {\ldots} + x_1 w^m=
\sum_{\a=1}^{m+1}x_\a w^{m+1-\a}$. We take $z \sim x_{m+1}$ or
$w \ll 1$. It follows that
\be
W^{\pr} dz = - \frac{1}{w^{m+1} } dw, \;\;\;\;\;\;
\frac{\pa W}{ \pa x_\a} = W^{\pr} \frac{\pa z}{ \pa x_\a} 
= W^{\pr} w^{m+1-\a}
\ee
Consequently:
\br
&&g (\frac{\pa W}{ \pa x_\a} , \frac{\pa W}{ \pa x_\b})= 
 - {\rm Res}_{z=\infty}\( \frac{(\pa W / \pa x_\a) (\pa W /\pa x_\b)}
{W^{\pr}}\) dz \lab{metrx} \\
&=& 
- {\rm Res}_{z=\infty} \( W^{\pr} w^{m+1-\a}w^{m+1-\b}\) dz 
= 
{\rm Res}_{w=\infty} \(\frac{w^{m+1-\a}w^{m+1-\b}}{w^{m+1} }\) dw=
 \d_{\a+\b=m+2} 
\nonu 
\er
Hence $x_\a$ are flat coordinates with the metric $\eta_{\a\b}=
\d_{\a+\b=m+2} $. The coefficients $v_j,\,j=1,{\ldots}, 
m+1$ of $W(z)$ are given in terms of the flat coordinates as 
\ct{aoyama}:
\br
v_k &=& \sum_{\a_1+{\ldots} +\a_k=(k-1)m+k} x_{\a_1}x_{\a_2}\cdots
x_{\a_k}, \quad k=1,{\ldots} ,m 
\lab{vkasx}\\
v_{m+1} &=& x_{m+1} \nonu
\er
Examples are :
\be
v_m = (x_m)^m, \; v_{m-1} = (m-1) x_{m-1}(x_m)^{m-1}, \,
{\ldots}, \, v_1=x_1
\lab{vkasxe}
\ee
To represent the remaining coefficients of $a_i, i=1,{\ldots} ,n$ 
of $W$ in terms of the flat coordinates we consider
a relation:
\be
z = w+ \frac{{\ti x}_1}{w} + \frac{{\ti x}_2}{w^2}+{\ldots} +
\frac{{\ti x}_n}{w^n}
\lab{anxti}
\ee
valid for large $z$ and $w \gg 1$. In this limit we impose a relation
$W= w^{n+1}/(n+1)$ from which it follows that
\be
W^{\pr} dz= w^n dw
, \;\;\quad 
\frac{\pa W}{ \pa {\ti x}_\g} = W^{\pr} \frac{\pa z}{ \pa {\ti x}_\g} 
= W^{\pr} w^{-\g}
\lab{Wprwn}
\ee
We find 
\br
&&g (\frac{\pa W}{ \pa {\ti x}_\g} , \frac{\pa W}{ \pa {\ti x}_\d})= 
{\rm Res}_{z\in {\rm Ker} W}\( \frac{(\pa W / \pa {\ti x}_\g )
(\pa W /\pa {\ti x}_\d)}
{W^{\pr}}\) dz  
\lab{metrxa}\\
&=&  {\rm Res}_{z \in {\rm Ker} W} 
\( W^{\pr} w^{-\g}w^{-\d}\) dz =
{\rm Res}_{w=0} w^{n-\g-\d} dw=
 \d_{\g+\d=n+1} 
\nonu
\er
Hence ${\ti x}_\g$ are flat coordinates with the metric 
${\ti \eta}_{\g\d}=
\d_{\g+\d=n+1} $.
By similar considerations $\eta_{\a \g}=0$ for $\a=1,{\ldots} ,m+1, \g =1,{\ldots}
,n$.

{}From expression \rf{anxti} and $W(z) = w^{n+1}/(n+1)$ one can find
relations between coefficients $a_\g$ and ${\ti x}_\g$ \ct{aoyama}
starting with $a_{n-1} = -{\ti x}_1$ and so on.

We will now show how to associate to the rational potentials $W$ 
canonical coordinates $u_i, i=1,{\ldots} ,n+m+1$ for which
the metric \rf{wmetric} becomes a Darboux-Egoroff metric.

Let $\a_i$, $i=1,{\ldots} ,n+m+1$ be roots of the rational potential 
$W(z)$ in \rf{wprime}. Equivalently,
$W^{\pr}(\a_i) =0$ for all $i=1,{\ldots} ,n+m+1$.
Thus $W^{\pr} (z)$ can be rewritten as 
\be
W^{\pr} (z) = \frac{\prod_{j=1}^{n+m+1}(z-\a_j)}{(z-v_{m+1})^{m+1}}
\lab{wprima}
\ee
Next, define the canonical coordinates as 
\be
u_i = W (\a_i) , \quad i=1,{\ldots} ,n+m+1
\lab{cancoord}
\ee
The identity :
\br
\d^i_j &=& \frac{\pa u_i}{\pa u_j} =  \frac{\pa  W (\a_i)}{\pa u_j} 
\nonu \\
&=& W^{\pr} (\a_i) \frac{\pa \a_i}{\pa u_j} + \frac{\pa  W }{\pa u_j} (\a_i)
= \frac{\pa  W }{\pa u_j} (\a_i)
\lab{wuiuj}
\er
implies that 
\be
\frac{\pa  W }{\pa u_j} (z) =  
\frac{\pa a_{n-1}}{\pa u_j} z^{n-1}+ {\ldots} +
\frac{\pa a_{0}}{\pa u_j}
+ \frac{\pa v_1/ \pa u_j}{z-v_{m+1}}+ {\ldots} 
+
\frac{v_m}{(z-v_{m+1})^{m+1}} \frac{\pa v_{m+1}}{\pa u_j}
\lab{wuja}
\ee
can be rewritten as 
\be
\frac{\pa  W }{\pa u_j} (z) =  
\frac{\prod_{k=1,j\ne k}^{n+m+1}(z-\a_k)}{(z-v_{m+1})^{m+1}}
\, \frac{(\a_j-v_{m+1})^{m+1}}{\prod_{k=1,j\ne k}^{n+m+1}(\a_j-\a_k)}
\lab{wujb}
\ee
Consider
\be
g ( \frac{\pa  W }{\pa u_i}, \frac{\pa  W }{\pa u_j}) = 
{\rm Res}_{z \in {\rm Ker} W^{\pr}} 
\( \frac{ ({\pa  W }/{\pa u_i})({\pa  W }{\pa u_j})}{W^{\pr}} \)dz
\lab{wmetuu}
\ee
Recalling \rf{wprima} and \rf{wujb} we find that 
$g ( {\pa  W }/{\pa u_i}, {\pa  W }/{\pa u_j})=0$ for $i\ne j$.
For $i=j$, we find
\br
g ( \frac{\pa  W }{\pa u_i}, \frac{\pa  W }{\pa u_i}) &= &
{\rm Res}_{z \in {\rm Ker} W^{\pr}} 
\( \frac{( {\pa  W }/{\pa u_i})^2}{W^{\pr}} \)dz \nonu \\
&=& 
\frac{(\a_i-v_{m+1})^{m+1}}{\prod_{j=1,j\ne i}^{n+m+1}(\a_i-\a_j)}
= \frac{ \pa a_{n-1}}{\pa u_i}
\lab{wmetuii}
\er
where the last identity was obtained by comparing coefficients of
the $z^{n-1}$ term in \rf{wuja} and \rf{wujb}.

Hence, in terms of the coordinates $u_i$ the metric can be rewritten
as $ g = \sum_{i=1}^N h_i^2(u) (d u_i)^2$ with the Lam\'e
coefficients :
\be
 h_i^2(u) = \frac{ \pa a_{n-1}}{\pa u_i}
\lab{lamea}
\ee
The fact that $h_i^2(u)$ is a gradient ensures that the rotation
coefficients $\b_{ij}$ are symmetric and therefore the metric becomes
the Darboux-Egoroff metric when expressed in terms of the 
orthogonal 
curvilinear coordinates $u_i$.

\sect{N=3 Models, Examples of Rational Landau - Ginsburg models }
\label{examples}
\subsection{n=m=1 model}
Consider the model with $n=m=1$ :
\be
W (z) = \h z^2 +x_1 +\frac{x_2}{z-x_3}
\lab{oldmodel}
\ee
where as coefficients we used the flat coordinates $x_1 =-{\ti x}_1$
and $x_2,x_3$ corresponding to $x_1,x_2$ of the previous section.
The flat coordinates $x_\a , \a=1,2,3$ 
are related to the flat metric :
\be
\eta^{\a\b} = \eta_{\a\b} = {\rm Res}_{z \in {\rm Ker} W^{\pr}}
\( \frac{ (\pa W /\pa x_\a)( \pa W /\pa x_\b) }{W^{\pr}} \)dz
= \threemat{1}{0}{0}{0}{0}{1}{0}{1}{0}
\lab{wmetric3}
\ee
The metric tensor can be derived from the more general expression involving
the structure constants
\be
c^{\a\b\g} = {\rm Res}_{z \in {\rm Ker} W^{\pr}}
\( \frac{ (\pa W /\pa x_\a)( \pa W /\pa x_\b)( \pa W /\pa x_\g) }{W^{\pr}} \)dz
\lab{strco}
\ee
through relation $\eta^{\a\b}= c^{\a\b1}$.
The non-zero values of the components of $c_{\a\b\g} $ are found from 
\rf{strco} to be :
\be
c^{111} = 1 ,\; c^{123}=1 , c^{222}= 1/x_2, \; c^{233}=x_3
, \; c^{333}=x_2
\ee
the other values can be derived using that $c^{\a\b\g} $ is symmetric in all
three indices.
These values can be reproduced from the formula \rf{cfxx}
with the prepotential :
\be
F (x_1, x_2, x_3) = \frac{1}{6} x_2 (x_3)^3 +  
\frac{1}{6} (x_1)^3 + x_1  x_2 x_3 +
\h (x_2)^2 \( \log x_2 -  \frac{3}{2}\)
\lab{prepo3}
\ee

The prepotential satisfies the quasi-homogeneity relation 
\rf{quasi} with $d_F=3$ with respect to the 
Euler vectorfield :
\be
E= x_1 \frac{\pa }{\pa x_1} + \frac{3}{2} x_2 \frac{\pa }{\pa x_2} +
\frac{1}{2}x_3 \frac{\pa }{\pa x_3} 
= x^1 \frac{\pa }{\pa x^1} + \frac{1}{2} x^2 \frac{\pa }{\pa x^2} +
\frac{3}{2}x^3 \frac{\pa }{\pa x^3} 
\lab{euler3}
\ee
We now adopt a general discussion of canonical coordinates to the case
$n=m=1$. 
Let $\a_i$, $i=1,2,3$ be roots of the polynomial 
\rf{wprime}, which at present case is 
$W^{\pr} (z) = z -{x_2}/{(z-x_3)^2}$.
So, $\a_i$ satisfy $W^{\pr}(\a_i) =0$ or $\a_i (\a_i -x_3)^2-x_2=0$
for all $i=1,2 ,3$.

Then, it follows by taking derivatives of $\a_i (\a_i -x_3)^2=x_2$
with respect to $x_2,x_3$ that 
\be
\frac{\pa \a_i}{\pa x_3} = \frac{2 \a_i}{3\a_i-x_3},\;\;\qquad 
\frac{\pa \a_i}{\pa x_2} = \frac{1}{(\a_i-x_3)(3\a_i-x_3)}
\lab{alpix}
\ee
and further that
\be
\frac{\pa u_i}{\pa x_3} = \frac{x_2 }{(\a_i-x_3)^2}=\a_i ,\;\;\qquad 
\frac{\pa u_i}{\pa x_2} = \frac{1}{\a_i-x_3}
\lab{uix}
\ee
for the canonical coordinates $u_i = W(\a_i)=\h \a_i^2 + x_1 + x_2/(\a_i-x_3)$.
We now present a method of inverting the derivatives in \rf{uix} or
alternatively to find the matrix elements $m_{ij}$ of the matrix 
$M$ from relation \rf{mdef}.
The sum of the canonical coordinates is equal to 
$\sum_{i=1}^3 u_i = 3 x_1 + x_3^2$ and therefore 
\be
1 = 3 \frac{\pa x_1}{\pa u_i} + 2 x_3 \frac{\pa x_3}{\pa u_i}
=h_i^2 \(3+ 2 x_3 \frac{\pa u_i}{\pa x_2}\)
\lab{1himi3}
\ee
where we used the fact that
\be
\frac{\pa x_3}{\pa u_i} = m^2_{i1} \frac{\pa u_i}{\pa x_2}
\lab{x3uia}
\ee
because of
\be
\frac{\pa x_\a}{\pa u_i} = m_{i1} m_{i\a} , \quad
\frac{\pa u_i}{\pa x_\a} = \eta_{\a \b} \frac{m_{i\b} }{m_{i1}}, \;\;\; 
h_i^2 = m^2_{i1}
\ee
Hence, from relation \rf{1himi3} it holds that $h_i^2=  \(3+ 2 x_3 \frac{\pa u_i}{\pa x_2}\)^{-1}$
or by using equation \rf{uix} that 
\be
\frac{\pa x_1}{\pa u_i} = h_i^2=  \frac{\a_i-x_3}{3\a_i-x_3}
\lab{x1ui}
\ee
Plugging the last equation into equation \rf{x3uia} and using relation
\rf{uix} we obtain
\be
\frac{\pa x_3}{\pa u_i} = \frac{1}{3\a_i-x_3}
\lab{x3uib}
\ee
Similarly, from 
\be
\frac{\pa x_2}{\pa u_i} = m^2_{i1} \frac{\pa u_i}{\pa x_3}
\lab{x2uia}
\ee
we obtain
\be
\frac{\pa x_2}{\pa u_i} = \frac{x_2}{(\a_i-x_3)(3\a_i-x_3)}=
 \frac{\a_i(\a_i-x_3)}{(3\a_i-x_3)}
\lab{x2uib}
\ee
Furthermore,
\be
\frac{\pa \a_i}{\pa u_j}= \frac{\pa \a_i}{\pa x_2}\frac{\pa x_2}{\pa u_j}
+\frac{\pa \a_i}{\pa x_3}\frac{\pa x_3}{\pa u_j}
\ee
gives for $i\ne j$:
\be
\frac{\pa \a_i}{\pa u_j}= \frac{1}{(3\a_i-x_3)(3\a_j-x_3)}\(
\frac{\a_j(\a_j-x_3)}{(3\a_i-x_3)}+2 \a_i\)
\lab{paapauj}
\ee
and for $i = j$ :
\be
\frac{\pa \a_i}{\pa u_i}= \frac{3\a_i}{(3\a_i-x_3)^2}
\lab{paapaui}
\ee
Using \rf{x1ui} and \rf{paapauj} we find
the rotation coefficients defined in \rf{rotco}
to be
\be
\b_{ij} = - \frac{(\a_k-x_3)(3\a_k-x_3)}{(3\a_i-x_3)(3\a_j-x_3)}
\frac{1}{\sqrt{(\a_i-x_3)(3\a_i-x_3)(\a_j-x_3)(3\a_j-x_3)}}
\lab{bmodel}
\ee
Its square is then
\be
\b_{ij}^2 = - \frac{1}{(\a_i-\a_j)^2}\frac{1}{(4x_3-3\a_k)^2}
\frac{\pa x_1}{\pa u_k}, 
\lab{bsmodel}
\ee
where $i,j,k$ are cyclic.
Recall that in equation \rf{vome} we have introduced the
functions $\om_k = (u_j-u_i) \b_{ij}$, where again we used the 
cyclic indices $i,j,k$.
The difference of canonical coordinates can be written as :
$u_j-u_i=(\a_i-\a_j)(3\a_k-4x_3)/2$ which together with equation \rf{bmodel}
yields:
\be
\om_k^2 =- \frac{1}{4} h_k^2 =- \frac{1}{4}\frac{\pa x_1}{\pa u_k}
=- \frac{1}{4} \frac{\a_k-x_3}{3\a_k-x_3}
\lab{omksmodel}
\ee
Since $ I=\sum_{i=1}^3 \pa / \pa u_i= \pa / \pa x_1$
then 
\be
\sum_{k=1}^3 \om_k = - \frac{1}{4}, \quad 
E (\log \tau) = \frac{1}{4}
\lab{etaurmd}
\ee
The explicit form of the roots $\a_i$ is needed to find
expressions for $\om_k$ and its dependence on the parameter
$s$.
It is convenient to introduce $q= x_2/(x_3)^3 $ and
$a_i = \a_i/x_3$ which 
satisfy equation $a_i(a_i-1)^2=q$.
Let us furthermore introduce a parameter $ \om$ such that
$q =4 (\om^2-1)^2 /(\om^2+3)^3$. This parametrization makes
it possible to obtain the compact expressions for $\om_k$.
The three solutions to the algebraic equation 
\be
a(a-1)^2 =q= 4 \frac{(\om^2-1)^2 }{(\om^2+3)^3}
\lab{paraeq}
\ee
are:
\be
a_1 = \frac{4 }{\om^2+3}, \;\; a_2 = \frac{(\om+1)^2 }{\om^2+3}, \;\; 
a_3 = \frac{(\om-1)^2 }{\om^2+3}
\lab{aroots}
\ee
Note, that $a_2 \leftrightarrow a_3$ under $\om \leftrightarrow -\om$
transformation, which shows that $\om$ is a purely imaginary 
variable.
First, we find that the variable $s$ from \rf{sdef} can be expressed as :
\be
s= \frac{(a_2-a_1)}{(a_3-a_1)}\frac{(3a_3-4)}{(3a_2-4)}=
\frac{(\om-3)^3(\om+1)}{(\om+3)^3(\om-1)}
\lab{sdefmodel}
\ee
Next, from relations $h_i^2= (a_i-1)/(3a_i-1)$ and equation  
\rf{omksmodel} we derive :
\be
\om_1^2 = - \frac{1}{4}\frac{(\om^2-1) }{(\om^2-9)}, \;\;\;
\om_2^2 =  \frac{1}{4}\frac{(\om+1) }{\om(\om-3)}, \;\;\;
\om_3^2 = - \frac{1}{4}\frac{(\om-1) }{\om(\om+3)}
\lab{ompara}
\ee
They provide solutions to the Euler top equations 
\rf{euta}-\rf{eutc}.
The corresponding function \ct{segert1,segert2}
\be
y(\om) = \frac{( \om -3 )^2({\om} +1  ) }{( \om +3 )({\om^2}+3 )}
\lab{painsol}
\ee
connected with $\om_k$'s through relations \ct{H1,H2,H3}:
\br
\om^2_1&=&-\frac{(y-s)y^2(y-1)}{s}\left(v-\frac{1}{2(y-s)}\right)\left(
v-\frac{1}{2(y-1)}\right)~,\nonu \\
\om^2_2 &=&
\frac{(y-s)^2y(y-1)}{s(1-s)}\left(v-\frac{1}{2(y-1)}\right)\left(
v-\frac{1}{2y}\right)~,\nonu \\
\om^2_3&=&-\frac{(y-s)y(y-1)^2}{(1-s)}\left(v-\frac{1}{2y}\right)\left(
v-\frac{1}{2(y-s)}\right) 
\lab{omtoy}
\er
with the auxiliary variable $v$ defined by equation
\be
\dder{y}{s}=\frac{y(y-1)(y-s)}{s(s-1)}\left(2v-\frac{1}{2y}-\frac{1}{2(y-1)}
+\frac{1}{2(y-s)}\right)
\lab{auxi}
\ee
is a solution of the Painlev\'e VI  equation  \ct{H1,H2,H3}:
\br
\frac{d^2 y}{d s^2} &=&\frac{1}{2}\left( \frac{1}{y} 
+\frac{1}{y-1}+\frac{1}{y-s}\right)
(\dder{y}{s})^2- \left( \frac{1}{s} +\frac{1}{s-1}+\frac{1}{y-s}\right)
\dder{y}{s}\nonu \\
&  +& \frac{y(y-1)(y-s)}{s^2(s-1)^2}\left[
\frac{1}{8} -\frac{s}{8y^2}+\frac{s-1}{8(y-1)^2}
+\frac{3s(s-1)}{8(y-s)^2}\right] \, ,
\lab{pain6}
\er
Introducing parameter $x=(\om-3)/(\om+3)$ one can rewrite expressions
\rf{painsol} and  \rf{sdefmodel} as :
\be
y = \frac{x^2(x+2)}{x^2+x+1}\; , \qquad s= \frac{x^3(x+2) }{2x+1}\, ,
\lab{cnh2}
\ee
which reproduces the $k=3$ Poncelet 
polygon solution of Hitchin \ct{H2,H3}.
Note, also that taking an inverse of solution \rf{painsol} and letting $\om
\to -\om$ produces the $k=6$ Poncelet 
polygon solution of Hitchin \ct{H2,H3}:
\be
y^{-1} (- \om ) = \frac{( \om -3 )({\om}^2 +3  ) }{( \om -1 )
({\om}+3 )^2} = \frac{x(x^2+x+1)}{2x+1}\, .
\lab{painsolk6}
\ee

We now proceed to calculate the underlying $\t$-function.
Our knowledge of the $\t$-function is based on equation 
\rf{tauvjv} from which we derive
that 
\be
\pa_j \log \t =  \sum_{i=1}^3 \b_{ij}^2 (u_i-u_j)\,.
\lab{pajltbu}
\ee
The identity $I (\log \t) =0$, shows that $\t=\t(x_2,x_3)$
is a function of two variables $x_2,x_3$. Furthermore, it
satisfies :
\be
E (\log \t)= \( \frac{3}{2} x_2 \frac{\pa }{\pa x_2} +
\frac{1}{2}x_3 \frac{\pa }{\pa x_3}\) \log \t= \frac{1}{4}
\lab{elogtmod}
\ee
A solution to the above equation is 
\be
\log \t= \frac{1}{4} \(\frac{1}{3} \log x_2 + \log x_3\) +
f\(\frac{1}{3} \log x_2 - \log x_3\) 
\lab{decot}
\ee
where $f(\cdot)$ is an arbitrary function of it's argument. 
In order to determine the function $f$ we use equation \rf{pajltbu}
to calculate the derivative
\be
\frac{\pa \log \t}{\pa x_3} = \sum_{j=1}^3 \frac{\pa u_j}{\pa x_3}
\pa_j \log \t = \sum_{i,j=1}^3 \a_j \b_{ij}^2 (u_i-u_j)
\lab{pajx3lt}
\ee
A calculation based on equation \rf{bsmodel} yields:
\be
 x_3 \frac{\pa }{\pa x_3} \log \t=  \frac{1}{8} \frac{1}{1-\frac{27}{4}q}
= \frac{1}{4} -
f^{\pr} \(\frac{1}{3} \log x_2 - \log x_3\) 
\lab{27q}
\ee
where the last equality was obtained by 
comparing with equation \rf{decot} (recall that $q= x_2/(x_3)^3 $).
Integration gives (ignoring an inessential integration constant) :
\be
f \(\frac{1}{3} \log x_2 - \log x_3\) =
\frac{1}{24} \( \log q + \log (-4 +27 q)\) 
\lab{fqres}
\ee
Using that $x_2= q x_3^3$ we can now rewrite $\log \t$
as
\be
\log \t = \frac{1}{4} \log x_3^2 + \frac{1}{24}  \log \(q^3(-4 +27 q)\) 
\lab{ltresq}
\ee
Inserting parametrization of $q$ from \rf{paraeq} 
and using relation $u_2-u_3=8 x_3^2 \om^3(\om^2+3)^{-2}$
we obtain the following expression for $\log \t$ :
\be
\log \t = \log (u_2-u_3)^{\frac{1}{4}} + \frac{1}{24}  
\log \((\om-1)^6 (\om+1)^6 (\om-3)^2 (\om+3)^2 \om^{-16}\)
\lab{ltresom}
\ee
It is easy to confirm $I (\log \t)=0$ and $E (\log \t)=1/4$
based on this expression.

\subsection{n=0,m=2 model}
Consider the model with $n=0, m=2$ in \rf{ratW} :
\be
W (z) =  z +\frac{x_1}{z-x_3}+\frac{x_2^2}{2(z-x_3)^2}
\lab{newmodel}
\ee
The flat coordinates $x_\a , \a=1,2,3$ 
are related to the flat anti-diagonal metric :
\be
\eta^{\a\b} = \eta_{\a\b} = {\rm Res}_{z \in {\rm Ker} W^{\pr}}
\( \frac{ (\pa W /\pa x_\a)( \pa W /\pa x_\b) }{W^{\pr}} \)dz
= \threemat{0}{0}{1}{0}{1}{0}{1}{0}{0}
\lab{newwmetric3}
\ee
The model is characterized by the prepotential
\be
F = \h x_3^2 x_1 +\h x_2^2 x_3 + \h x_1^2 \log (x_2)
\lab{nmprepot}
\ee
which generates the structure constants according to relation
\rf{cfxx} and possesses homogeneity $d_f=4$ with respect to
the Euler operator
\be
E= 2x_1 \frac{\pa }{\pa x_1} + \frac{3}{2} x_2 \frac{\pa }{\pa x_2} +
x_3 \frac{\pa }{\pa x_3} 
= x^1 \frac{\pa }{\pa x^1} + \frac{3}{2} x^2 \frac{\pa }{\pa x^2} +
2 x^3 \frac{\pa }{\pa x^3}
\lab{nmeuler3}
\ee
Plugging the values $d_1=1,d_2=3/2,d_3=2,d_F=4$ into the 
relation \rf{randdf} we find that the homogeneity of
$\log \t$ is again equal to $R^2 =1/4$.
The roots $\a_i$, $i=1,2,3$ of $W^{\pr} (\a_i)=0$
with $W$ given in \rf{newmodel}
satisfy equation:
\be
(\a-x_3)^3 - (\a-x_3)x_1-x_2^2=0
\lab{nmeqa}
\ee
It is convenient to introduce instead variables $f_i = \a_i -x_3$
which satisfy
\be
f^3 - f x_1-x_2^2=0
\lab{nmeqf}
\ee
Clearly, $f_i$ are functions of $x_1$ an $x_2$ only 
and by taking derivatives of \rf{nmeqf} we obtain:
\be
\frac{\pa f_i}{\pa x_1} = \frac{f_i}{3f_i^2-x_1},\qquad 
\frac{\pa f_i}{\pa x_2} = \frac{2x_2}{3f_i^2-x_1)}
\lab{fix}
\ee
In terms of variables $f_i$ the canonical coordinates
become
\be
u_i = W(\a_i)= x_3 + \frac{3}{2} f_i + \h \frac{x_1}{f_i}
\lab{uifinm}
\ee
and satisfy $\sum_{i=1}^3 u_i = 3 x_3 - x_1^2/2x_2^2$.
We find that the Lam\'e coefficients are given by:
\be
h_i^2 = \pder{x_1}{u_i} = \frac{f_i^3}{3f_i^2-x_1}
\lab{lamnm}
\ee
The homogeneity of $h_i^2$ is found after applying a general formula
$E \pa_i = \pa_i E -\pa_i$ to the above equality with the result
\be
E (h_i^2) = E (\pder{x_1}{u_i}) = \pa_i E (x_1) - \pa_i x_1
= (d_3-1)\pder{x_1}{u_i}= (d_3-1) h_i^2 = h_i^2\, .
\lab{ehid1}
\ee
Here we used that $d_3= \sum_{\a=1}^3 d_\a \eta^{\a 1} =  2$ for the 
model under consideration. 
Thus the current Lam\'e coefficients $h_i^2$ will not 
depend on only one variable $s$.
Notice, that
in the  previous $n=m=1$ model the value $d_1= \sum_{\a=1}^3 d_\a \eta^{\a 1}= 1$ was 
consistent with homogeneity of
$h_i^2$ being zero. 

The corresponding rotation coefficients 
\be
\b_{ij}= \1{2\sqrt{\pder{x_1}{u_i}\pder{x_1}{u_j}}}
\frac{\pa^2 x_1}{\pa u_i \pa u_j}
\lab{bijnm}
\ee
are calculated straightforwardly from knowledge of \rf{lamnm}
and following derivatives
\be
\pder{f_i}{u_j} = \frac{2 x_2^2 f_j +f_j^3f_i}{(3f_i^2-x_1)(3f_j^2-x_1)} ,
\; \;\; i \ne j
\lab{fiuj}
\ee
They are explicitly given by 
\be
\b_{ij} = \1{\sqrt{f_if_j}} \1{\sqrt{3f_k^2-4x_1}}
\frac{N_k}{(f_i-f_j)^2 (3f_k^2-x_1)^{(5/2)}} \lab{bijnma}
\ee
where $N_k=8x_2^4f_k^2-4x_1x_2^4+f^4_if^4_j$ with indices $i,j,k$
being cyclic.

The result for $ \om_k= (u_j-u_i) \b_{ij}$ is
\be
\om_k^2  = \frac{r}{4} \frac{3g_k^2-4}
{(g_k-3r) (3g_k^2-1)^{5}} \( 8 g_k^4-4g_k^2+\frac{r^2}{g_k^2}\)^2
\lab{omknm}
\ee
where $g_k, k=1,2,3$ are three roots of the equation $g^3-g-r=0$
with $r=x_2^2/x_1^{3/2}$.

One verifies that $\om_k^2 $ from \rf{omknm} do indeed satisfy 
$ \sum_{k=1}^3 \om_k^2 =- 1/4$ and the Euler top equations
\rf{euta}-\rf{eutc}.
This model provides an example of the Lam\'e coefficients
with non-zero homogeneity and consequently $\om_k^2$ are not
proportional to the Lam\'e coefficients.

Is it possible though to make another choice for solutions
of the Euler top equation in this model.
Consider, namely
\be
{\ti h}_i^2 = \pder{x_3}{u_i} = \frac{f_i^2}{3f_i^2-x_1}
\lab{lamnmti}
\ee
with the properties $\sum_{i=1}^3 {\ti h}_i^2 =1$ and
$ E ({\ti h}_i^2 ) = (d_1-1) {\ti h}_i^2 =0$

The corresponding rotation coefficients
\be
{\ti \b}_{ij}= \1{2\sqrt{\pder{x_3}{u_i}\pder{x_3}{u_j}}}
\frac{\pa^2 x_3}{\pa u_i \pa u_j}
\lab{bijnm3}
\ee
are found to be given by
\be
{\ti \b}_{ij}= \frac{x_2^2}{2} 
\frac{3f_k^2-x_1}{(3f_i^2-x_1)^{(3/2)}(3f_j^2-x_1)^{(3/2)}} \lab{bijnmti}
\ee
which produces ${\ti \om}_k = (u_j-u_i) {\ti \b}_{ij}$ 
equal to
\be
{\ti \om}_k^2 = - \1{16} {\ti h}_k^2= 
- \1{16} \frac{f_k^2}{(3f_k^2-x_1)}
\lab{newtiom}
\ee
which satisfy 
$ \sum_{k=1}^3 \om_k^2 =- 1/16$ and the Euler top equations
\rf{euta}-\rf{eutc}.

\sect{Discussion}
\label{outlook}
This paper shows how to derive  the canonical integrable structures
for some class of rational Lax functions associated
with a particular reduction of the dispersionless 
KP hierarchy. This derivation generalizes the well-known
construction of the monic polynomials \ct{Du-lectures,Hlectures}.
The three-dimensional examples provide solutions to the
Painlev\'e VI equation. 
Given that the flows of the canonical integrable models
can essentially be reformulated as isomonodromic deformations its connection to the
sixth Painlev\'e equation is not surprising.

Much of the discussion is centered around the tau
functions from which all the objects of the Darboux-Egoroff metric
can be derived.
The tau functions of the three-dimensional examples had
a scaling dimension of $R^2=1/4$ and the corresponding prepotentials 
contained logarithmic terms.
For the scaling dimensions, $R^2=n^2$ such that $n$ is an integer,
the multi-component KP hierarchy provides a framework
for the construction of canonical integrable hierarchies.
The longterm goal of this work is to search for a universal 
approach 
to the formulation of the canonical integrable models 
which would include models with fractional scaling dimensions
as the ones encountered in examples based on the rational 
potentials of the LG type.

Further progress is needed for the classification
of relevant rational reductions of
the Toda and KP hierarchies and
the canonical integrable models with associated Darboux-Egoroff metrics
which can be derived from related rational Lax functions.
This problem is currently under investigation.

\vskip 10pt \noindent
{\bf Acknowledgments} \\
H.A. was partially supported by FAPESP and NSF (PHY-9820663).
A.H.Z. and J.F.G were partially supported by CNPq.
The authors thank Johan van de Leur for reading the manuscript and his comments
on it.

\end{document}